\documentclass[a4paper,11pt,floatfix,showpacs,tightenlines,showkeys,superscriptaddress,amsmath,amssymb,nofootinbib]{article}
\usepackage{jcappub}
\usepackage{amssymb,amsbsy,epsfig,color,graphicx}
\usepackage{color}
\usepackage{[longtable}
\usepackage{array}
\usepackage{dcolumn}   
\usepackage{cellspace}
\usepackage{mathtools}
\usepackage{amstext}
\usepackage{amssymb}
\usepackage{stmaryrd}
\usepackage{stackrel}
\usepackage{graphicx}
\usepackage{esint}
\usepackage[utf8]{inputenc}
\usepackage{blindtext}
\usepackage{float}
\restylefloat{table}
\usepackage{booktabs}
\usepackage{enumitem} 

\usepackage{etoolbox} 
\usepackage{lipsum} 
\usepackage[capitalize]{cleveref}

\usepackage{multirow}
\usepackage[caption=false]{subfig}
\renewcommand\[{\begin{equation}}
\renewcommand\]{\end{equation}}
\newcommand{\al}{\alpha}
\newcommand{\bt}{\beta}

\newcommand{\ba}{\begin{eqnarray}}
\newcommand{\ea}{\end{eqnarray}}

\makeatletter

\appto{\appendix}{%
	\@ifstar{\def\thesection{\unskip}\def\theequation@prefix{A.}}%
	{\def\thesection{\Alph {section}}}%
}
\makeatother

\begin{document}

	\title{Conformally-flat, non-singular static metric in infinite derivative gravity}
	
	\author[a,b,c]{Luca Buoninfante,}
	\author[d,e,f]{Alexey S. Koshelev,}
	\author[a,b]{Gaetano Lambiase,}
	\author[d]{Jo\~ao Marto,}
	\author[c,g]{Anupam Mazumdar}

	\affiliation[a]{Dipartimento di Fisica "E.R. Caianiello", Universit\`a di Salerno, I-84084 Fisciano (SA), Italy}
	\affiliation[b]{INFN - Sezione di Napoli, Gruppo collegato di Salerno, I-84084 Fisciano (SA), Italy}
	\affiliation[c]{Van Swinderen Institute, University of Groningen, 9747 AG, Groningen, The Netherlands}
	\affiliation[d]{Departamento de F\'isica and Centro de Matem\'atica e Aplica\c c\~oes, Universidade da Beira Interior, 6201-001 Covilh\~a, Portugal}
	\affiliation[e]{Theoretische Natuurkunde, Vrije Universiteit Brussel}
	\affiliation[f]{The International Solvay Institutes, Pleinlaan 2, B-1050, Brussels, Belgium}
	\affiliation[g]{Kapteyn Astronomical Institute, University of Groningen, 9700 AV, Groningen, The Netherlands}
	
	\date{\today}
	

\abstract{In Einstein's theory of general relativity the vacuum solution yields a blackhole with a curvature singularity, where there exists a point-like source with a Dirac delta distribution which is introduced as a boundary condition in the static case. It has been known for a while that {\it ghost-free} infinite derivative theory of gravity can ameliorate such a singularity at least at the level of linear perturbation around the Minkowski  background. In this paper, we will show that the Schwarzschild metric does not satisfy the boundary condition at the origin within infinite derivative theory of gravity, since a Dirac delta source is smeared out by non-local gravitational interaction. We will also show that the spacetime metric becomes conformally-flat and singularity-free within the non-local region, which can be also made devoid of an event horizon. Furthermore, the scale of non-locality ought to be as large as that of the Schwarzschild radius, in such a way that the gravitational potential in any metric has to be always bounded by one, implying that gravity remains weak from the infrared all the way up to the ultraviolet regime, in concurrence with the results obtained in [arXiv:1707.00273]. The singular Schwarzschild blackhole can now be potentially replaced by a non-singular compact object, whose core is governed by the mass and the {\it effective} scale of non-locality.}

\maketitle


\section{Introduction}

Since Einstein's work, the theory of general relativity (GR) has been subjected to many experimental tests, and  each one of them have shown an exemplary agreement between the observations and the theoretical predictions in the infrared (IR) regime, i.e. at large distances and late times \cite{-C.-M.}. Even the recent detection of gravitational waves emission due to the merging of binaries \cite{-B.-P.} have shown an excellent matching with the theory.

However, despite its great achievements, there are still open questions, which strongly suggest that GR is incomplete in the ultraviolet (UV) regime, for example cosmological and blackhole singularities at the classical level persist, and the non-renormalizability at the quantum level is still a hard problem to tackle. There is also an apparently simpler question that still needs to be answered - to what extent do we know the gravitational interaction at short distances? Indeed, the $1/r$-fall of the Newtonian potential has been tested only up to distances of the order of $10$ micrometers, in torsion balance experiments \cite{-D.-J.}. In terms of energy, it means that our knowledge of the gravitational interaction is limited {\it only} up to $0.01$eV (in Natural Units $\hbar=1=c$). The extrapolation of Einstein's gravity all the way up to the Planck scale, $M_p\sim 10^{19}$GeV,  in the abyss of more than $30$ order of magnitude, is indeed a mere speculation.

Over the past years, one of the most straightforward attempts to generalize the Einstein-Hilbert action was to introduce higher order derivatives through quadratic terms in the curvature, like $\mathcal{R}^2,\mathcal{R}_{\mu\nu}\mathcal{R}^{\mu\nu},\mathcal{R}_{\mu\nu\rho\sigma}\mathcal{R}^{\mu\nu\rho\sigma}$. In Ref.\cite{-K.-S.} it was shown that such a quadratic action turns out to be power-counting renormalizable in $4$ dimensions but, unfortunately, it is also happened to be a non-unitary theory, because of the presence of a spin-$2$ {\it ghost} as a dynamical degree of freedom. Such a conflict between the renormalizability and the unitarity seemed to be impossible to overcome, and was a crucial signal against the possibility to formulate a consistent theory of quantum gravity at the perturbative level. Of course, the aforementioned result is strongly based on {\it local} gravitational action, but what happens if we were to give up locality?

Recently, it has been noticed that the diatribe between renormalizability and unitarity can be potentially resolved by considering a non-local action, where non-locality is introduced through form-factors containing {\it infinite order covariant derivatives.} The form of the gravitational action was first conjectured in Refs. \cite{krasnikov,kuzmin,Tomboulis:1997gg,Tseytlin:1995uq,Siegel:2003vt,Biswas:2005qr}, but only in the last decade the full infinite derivative quadratic action has been derived in a more systematic way in Refs.\cite{Biswas:2011ar,Biswas:2016etb,Biswas:2016egy}, around constant curvature backgrounds.

The authors were also able to formulate a {\it ghost-free} condition in order to preserve the tree-level unitarity, emphasizing the need of {\it infinite covariant derivatives} in order to avoid ghost-like degrees of freedom in the physical massless, transverse and traceless graviton spectrum for Minkowski~\cite{Biswas:2011ar} and (anti-)de Sitter backgrounds~\cite{Biswas:2016etb,Biswas:2016egy}. In particular, they have also shown that the presence of infinite covariant derivative form-factors can also improve the short-distance behavior and solve problems like cosmological singularities \cite{Biswas:2005qr,Biswas:2010zk,Biswas:2012bp}, and singularity of the Newtonian potential \cite{Biswas:2011ar}. In Ref.\cite{Koshelev:2017bxd} it was argued that the presence of infinite order derivatives can resolve the singularity for astrophysical collapsing objects due to the non-local smearing of the space-time and of the weakening of the gravitational interaction at short distances. The dynamical avoidance of blackhole singularity (and an absence of a trapped surface) has been studied by Frolov and co-authors in Refs.~\cite{Frolov:2015bia,Frolov:2015usa}, and Frolov has also conjectured the possibility of a mass-gap in the context of {\it ghost-free} infinite derivative gravity~\cite{Frolov}. Indeed, we have been able to show that at the linear level, the Kretschmann invariant is non-singular and the metric in the non-local regime asymptotes to a conformally-flat in the static background~\cite{Buoninfante:2018xiw}.

Recently, a non-static, Kerr-type solution has also been constructed, but without a ring singularity in the {\it ghost-free} infinite derivative gravity~\cite{Cornell:2017irh}.  Furthermore, resolution of curvature singularity in membranes have also been studied~\cite{Boos:2018bxf}. At a quantum level infinite derivative form-factors can ameliorate the non-local behavior of the theory as pointed out in Refs. \cite{kuzmin,Tomboulis:1997gg,Talaganis:2014ida,Tomboulis:2015,Chin:2018puw}.

Given all these exciting results in this field of research, there have been attempts to address the full non-linear, non-local field equations for the {\it ghost-free} infinite derivative theory of gravity~\cite{Biswas:2013cha}. Only, this year two very interesting results have been obtained by~\cite{Koshelev:2018hpt}, and \cite{Koshelev:2018rau}. In the former paper, the authors have argued that the metric solution like Schwarzschild is not a vacuum solution for the infinite derivative gravity, and in the latter paper the Kasner type anisotropic metric has been shown to be not a solution for the full vacuum field equations of the infinite derivative gravity. 

It is worth mentioning here that, strictly speaking, the Schwarzschild metric solves Einstein's equations in the vacuum everywhere except at the origin $r=0$, where one has to impose a boundary condition by introducing a Dirac delta source, which is governed by the mass of the blackhole. The Birkhoff's theorem guarantees that a unique static and spherically symmetric metric solution can be obtained in this case. As we will show in this paper, we have a marked difference from the Schwarzschild geometry in the context of non-local, infinite derivative gravitational action.

The aim of this paper is to understand the space-time metric in the non-local region and show that the full field equations admit a regular metric solution which is conformally-flat and free from curvature singularity in the limit $r\rightarrow0$, as opposed to $1/r$ metric potential like in Schwarzschild's case. This matches with our earlier expectations of Refs.~\cite{Koshelev:2017bxd,Buoninfante:2018xiw}. In particular, we will show that in the infinite derivative theory, the boundary condition at $r=0$ cannot be satisfied. The infinite derivative gravity does not see a point-like source, it effectively sees a smeared system. In this respect the usual notion of vacuum solution which applies for the Schwarzschild metric in GR will not apply for the infinite derivative gravity. This smearing effect allows to have metric solutions whose functional form is regular at $r=0$; however, the presence of infinite derivatives is not sufficient in order to avoid curvature singularities. Indeed, we will show that also the ghost-free condition is crucial to select the right form of the metric potentials, thus preventing the presence of curvature singularities and making the spacetime conformally-flat in the non-local region.


\section{Infinite derivative gravity}

The most general quadratic action allowed by general covariance, parity-invariant and torsion-free, was constructed around maximally symmetric space-times in Refs. \cite{Biswas:2011ar,Biswas:2016etb}, and it is given by~\footnote{\label{foot:1} The original action was derived in terms of the Riemann tensor, so that one has
\begin{equation*}\label{act}
S= \displaystyle \frac{1}{16\pi G}\int d^4x\sqrt{-g}\left[\mathcal{R}+\alpha_c \left(\mathcal{R}F_1(\Box_s)\mathcal{R}+\mathcal{R}_{\mu\nu}F_2(\Box_s)\mathcal{R}^{\mu\nu}+\mathcal{R}_{\mu\nu\rho\sigma}F_{3}(\Box_s)\mathcal{R}^{\mu\nu\rho\sigma}\right)\right], 
\end{equation*}
where the new form-factor $F_i(\Box_s)$ are related to the others $\mathcal{F}_i(\Box_s)$ by the following relations:
\begin{equation*}
F_1(\Box_s)=\mathcal{F}_1(\Box_s)-\frac{1}{3}\mathcal{F}_3(\Box_s),\,\,\,\,\,\,\,\,\,\,\,\,F_2(\Box_s)=\mathcal{F}_1(\Box_s)-2\mathcal{F}_3(\Box_s),\,\,\,\,\,\,\,\,\,\,\,\,F_3(\Box_s)=\mathcal{F}_3(\Box_s).
\end{equation*}}
\begin{equation}
\begin{array}{rl}
S= & S_{EH}+S_{q}\\
= & \displaystyle \frac{1}{16\pi G}\int d^4x\sqrt{-g}\left[\mathcal{R}+\alpha_c\left(\mathcal{R}\mathcal{F}_1(\Box_s)\mathcal{R}+\mathcal{R}_{\mu\nu}\mathcal{F}_2(\Box_s)\mathcal{R}^{\mu\nu}+\mathcal{C}_{\mu\nu\rho\sigma}\mathcal{F}_{3}(\Box_s)\mathcal{C}^{\mu\nu\rho\sigma}\right)\right],\label{eq:1}
\end{array}
\end{equation}
where $S_{EH}$ is the Einstein-Hilbert action, and $S_{q}$ takes into account of all the quadratic terms in the curvature which would play a crucial role in the non-local regime; $\Box_s\equiv \Box/M_s^{2}$, where $\Box=g_{\mu\nu}\nabla^{\mu}\nabla^{\nu}$ is the d'Alambertian operator, and we adopt the mostly positive metric convention $(-,+,+,+)$. $G=1/M_p^{2}$ is Newton's constant, while the coupling constant $\alpha_c\sim 1/M_s^{2}$ is dimensionfull and it is related to the parameter $M_s$, which represents the scale of non-locality. The gravitational interactions become non-local in the region $r\leq 1/M_s$, and for momenta; $p^{\mu}p_{\mu}\geq M_s^2$. Indeed in the IR, we recover the Einstein-Hilbert action, for $\Box/M_s \rightarrow 0$. 

The gravitational form-factors $\mathcal{F}_i's$ are  reminiscence to any massless theory which has derivative interaction, and they are analytic function of $\Box_s$, which can be recast in terms of an infinite series:
\begin{equation}
\mathcal{F}_i(\Box_s)=\sum\limits_{n=0}^{\infty}f_{i,n}\Box_s^n.
\end{equation}
By analyzing the perturbative spectrum of the action in Eq.\eqref{eq:1} around maximally symmetric backgrounds, it turns out that in order to avoid tachyonic-like instabilities and ghost-like degrees of freedom, and thus to preserve the tree-level unitarity, one has to demand the form-factors $\mathcal{F}_i(\Box_s)$ to be expressed in terms of {\it exponential of an entire function} \cite{Siegel:2003vt,Biswas:2005qr,Biswas:2011ar,Biswas:2016etb,Biswas:2016egy}. This is due to the fact that {\it exponential of an entire function} does not have any zeroes in the finite complex plane, therefore no new poles arise in the graviton propagator, see for details in Ref.\cite{Biswas:2011ar}.

\section{Modifying the Schwarzschild geometry}\label{mod-sch}

\subsection{Non-local vs local action: blackhole as an Euclidean hole }
We will make a simple but potent argument to show when the term $S_q$ is relevant and dominates compared to the Einstein-Hilbert contribution, $S_{EH}$.  Since, we are dealing with a static geometry, let us introduce a characteristic length scale $L$, so that one has $dx\sim L$ and $\partial_x\sim1/L$; in the same way all curvature tensors will scale as $\mathcal{R}\sim 1/L^2.$ Thus, the two contributions in Eq.\eqref{eq:1} will scale as
\begin{equation}
S_{EH}\sim M_p^2\int d^4x\sqrt{-g}\mathcal{R}\sim M_p^2L^2,\,\,\,\,\,\,\,\,\,\,\,\,\,S_{q}\sim M_p^2 \int d^4x\sqrt{-g}~\alpha_c\left[\mathcal{R}\mathcal{F}_1(\Box_s)\mathcal{R}+\cdots \right]\sim \frac{M_p^2}{M_s^2}.\label{eq:2}
\end{equation}
Note that the quadratic curvature part of the action is scale invariant, and the {\it only} characteristic length scale is determined by $1/M_s$, where $M_s \ll M_p$. The action in the non-local regime virtually has no scales, which in a way suggests that any physical solution in this regime should be also scale invariant. The full gravitational action will be roughly given by:
\begin{equation}
S\sim M_p^2L^2 + \frac{M_p^2}{M_s^2}=M_p^2L^2\left(1+\frac{1}{M_s^2L^2}\right).\label{eq:3}
\end{equation}
Note that for characteristic scales larger than the scale of non-locality, $L\gg1/M_s$, the term $S_{q}$ is negligible, i.e. we are in the IR regime where the Einstein-Hilbert term dominates in the gravitational action, which indeed preserves all the good properties of GR in the IR. While in the limit $L\ll 1/M_s$, the infinite derivative quadratic curvature term is now no longer negligible, and dominates the full action, $S_{q}\gg S_{EH}$.\footnote{Interesting point to note here that if we had form factors $\mathcal{F}_i$, {\it purely} made up of finite derivatives, we would have a local theory with extra dynamical degrees of freedom. In particular, the coefficient in front the curvature squared, $\alpha_c$, would be related to the masses of the new degrees of freedom, i.e. a fixed number. In such a case there would be {\it no} analogue of the inequality in Eq.\eqref{eq:5}, namely that there would always be a critical value for the mass $m$ for which the inequality will not be satisfied, and the Einstein-Hilbert term would  come to dominate, yielding singular metric solution. 
	As an example we can consider Stelle's  quadratic curvature theory of gravity, for which singular metric solutions exist \cite{-K.-S.,Stelle}:
	\begin{equation}
	S=\frac{1}{16\pi G}\int d^4x\sqrt{-g} \left(\mathcal{R}+\alpha \mathcal{R}_{\mu\nu}\mathcal{R}^{\mu\nu}+\beta \mathcal{R}^2 \right),
	\end{equation}
	where other than the massless transverse spin-$2$ graviton, we also have two extra particles in the physical spectrum: a massive spin-$0$ with mass $m_0$ and a massive spin-$2$ ghost with mass $m_2$. By studying the pole structure of the propagator around the Minkowski background one can show that the two masses are related to the coefficients $\alpha$ and $\beta$ through the following relations:
	\begin{equation}
	m_0=(\alpha + \beta)^{-1/2},\,\,\,\,\,\,\,\,\,\,\,m_2=-(\alpha/2)^{-1/2}.
	\end{equation}
	It is now clear that the coefficient $\alpha$ and $\beta$ are fixed parameters in a local theory. The same happens for any local theory with finite order derivatives. Note also that, in order to avoid new degrees freedom, infinite derivatives are {\it not} sufficient, but we also need to constrain the three form-factors $\mathcal{F}_i$ in Eq.\eqref{eq:1} by demanding the transverse massless spin-$2$ graviton to be the only propagating degree of freedom; see also Subsection \ref{lin-sol}.\label{non-fixed}}

In the Einstein-Hilbert action the vacuum solution yields a Schwarzschild-like geometry with a Dirac delta source distribution at $r=0$, with a Schwarzschild horizon at $r_{sch}=2Gm$, where $m$ is the source mass of the blackhole. For $r > 2Gm$, the metric potential is $2|\phi(r)| =2Gm/r < 1$, and treated as a perturbative solution, while for $r< 2Gm$, the metric potential increases and blows up at $r=0$, where the Kretschmann invariant also shows a singular behavior. Definitely, the gravitational potential gets modified and becomes bigger than one for $r< 2GM$, in spite of the fact that the curvature remains tiny everywhere near the horizon, for astrophysical blackholes. In the Schwarzschild metric, indeed the horizon marks the demarcation between perturbative and non-perturbative aspects. So far there are no experimental evidences for the presence or the absence of an event horizon in astrophysical blackholes; the LIGO-VIRGO data cannot conclusively say the presence of an event horizon yet, see~\cite{Cardoso:2017cqb}.

Given all these facts, we would have to determine what should be the value of $L$ at which we would expect non-local interactions to dominate, i.e. $S_q\gg S_{EH}$?
From a physical point of view, if we demand that the relevant contribution due to $S_{q}$ becomes dominant at length scales of the order of the Schwarzschild radius, i.e. $L\sim 2Gm$, then the relation in Eq.\eqref{eq:3} becomes
\begin{equation}
S\sim\frac{4m^2}{M_p^2}\left(1+\frac{M_p^4}{4M_s^2m^2}\right)\,, \label{eq:4}
\end{equation}
therefore, when $S_{q}$ dominates over the Einstein-Hilbert term, then it yields the following relation\footnote{Note that the factor $2$ in $2/M_s$ is just a convention to be consistent with the factor $2$ entering in the modified gravitational potential; see Eq.\eqref{eq:8}.}:\footnote{Note that the presence of infinite derivatives in the gravitational action is a necessary condition in order to satisfy the inequality given by Eq.\eqref{eq:5}, since the scale of non-locality $M_s$ is {\it not} a fixed parameter; see also footnote \ref{non-fixed}.}
\begin{equation}
2mM_s<M_p^2 \Longleftrightarrow r_{\rm sch}< \frac{2}{M_s}\,.\label{eq:5}
\end{equation}
Physically, it means that at a characteristic scale $L\sim r_{\rm sch}$ the $S_q$ part of the action in Eq.\eqref{eq:1} plays a crucial role {\it only} if the non-local region engulfs the Schwarzschild radius, i.e. 
\begin{equation}
r_{NL}\sim 2/M_s> r_{sch}\,.
\end{equation}
This is indeed an interesting length scale, because  $r_{NL} \sim 2/M_s$  has appeared before {\it purely} from the linear considerations. In Ref.\cite{Biswas:2011ar}, it was shown that for $ r > r_{NL}\sim 2/M_s$, the gravitational metric potential would follow {\it strictly} GR's predictions, i.e. $1/r$-fall of the Newtonian potential, see this discussion below. Note that, in fact we can take $m\rightarrow \infty$ and $M_s\rightarrow 0$, and we would still satisfy the above bound Eq.(\ref{eq:5}). This means that we can consider supermassive astrophysical blackholes with  billion times solar mass, i.e. mass $m> 10^{6}\times 10^{33}\sim 10^{39}$ grams, and we can still be bounded by Eq.(\ref{eq:5}) as long as $M_s< 10^{-8}$eV~\footnote{Is it possible to realize such a small $M_s$? Is this scale $M_s$ a fixed number? The answer is no! This depends on the physical system and the interactions. In the case of blackholes, which can be seen as a bound-state made up of $N$ gravitons, the fundamental scale of non-locality can shift towards the IR, by $M_s \rightarrow M_s/\sqrt{N}$~\cite{Koshelev:2017bxd,Buoninfante:2018mre}.\label{shift}}. Indeed, if the photon is probing this region with momentum greater than $10^{-8}$eV, the photon interaction with the astrophysical object has to be dealt in the framework of non-local quantum field theory \cite{Buoninfante:2018mre}. The photon does not see a local graviton, and a conventional space-time. The details ought to be worked out and we leave some of these calculations for future studies. Furthermore, the above condition Eq.(\ref{eq:5}) also ensures that the energy density of the non-local, non-singular compact system with $r_{NL}\sim 2/M_s$ is always bounded below the Planckian energy density:
\begin{equation}
\rho\sim mM_s^3 < M_s^2M_p^2 < M_p^4\,,
\end{equation}
which is violated in Einstein's GR at $r=0$, as compared to infinite derivative gravity, where locality is lost within $r_{NL}\sim 2/M_s > r_{sch}$.\\

{\it Indeed, the above analysis does not assume any linearity of the action, it {\it merely} demands that at scales of the order of the Schwarzschild radius, i.e. $L\sim r_{\rm sch}=2Gm$, the infinite derivative part of the gravitational action will {\it only} dominate over the Einstein-Hilbert contribution, provided that the scale of non-locality is larger than that of the Schwarzschild radius, irrespective of the mass, $m$, of the source. Indeed, this is a non-perturbative statement, which we would expect to hold true in all regimes, both linear and non-linear. Moreover, within this region the energy density of the system is always bounded below the Planckian energy density. This is indeed one of the most desirable physical expectation in $4$ dimensions that we can never ever exceed the Planckian energy density in any kind of physical experiment, including the death of any star! }\\
 
Now this is an excellent juncture, where we can revert the question and ask: what non-local interactions ought to do with the weakening of the gravitational potential?
In fact, if we were to allow $S_{EH} > S_q$, we  would expect to have a region of space-time where $2|\Phi| > 1$, this is typical of the Einstein-Hilbert action, where one would expect formation of a trapped surface, apparent horizon, and an event horizon, see for a detailed discussion in Ref.\cite{Frolov-book}. However, one of the consequences of $S_q> S_{EH}$ is that it modifies the Schwarzschild geometry, the space-time structure becomes non-local and non-locality is such that it {\it weakens} the gravitational interaction. Inside this non-local region acausal effects may emerge such that the very concept of "clock" and "measuring stick" does not make sense, the usual notion of lightcone structure does not apply in the region $r<1/M_s$; see also Ref.\cite{Buoninfante:2018mre}. Indeed, instead of a conventional blackhole, we may now imagine a non-local object whose size is approximatively given by $\sim 2/M_s> r_{sch}$~\cite{Koshelev:2017bxd}, see Fig.1 for an illustration. However, such a non-local region can be well described with tools of Euclidean field theory: in this picture the blackhole becomes {\it effectively} an {\it Euclidean hole}. This also helps ameliorating the blackhole-information loss paradox, in a very similar fashion as that of the fuzz-ball scenario, for a review see~\cite{Mathur:2005zp}~\footnote{ It will be a bit premature to draw the direct comparison with the fuzz-ball scenario. We would require some more detailed work to establish this connection. However, non-local interaction is an inherent feature of string field theory, for a review see~\cite{Siegel:1988yz}.}.
 \begin{figure}[t!]
 	\includegraphics[scale=0.30]{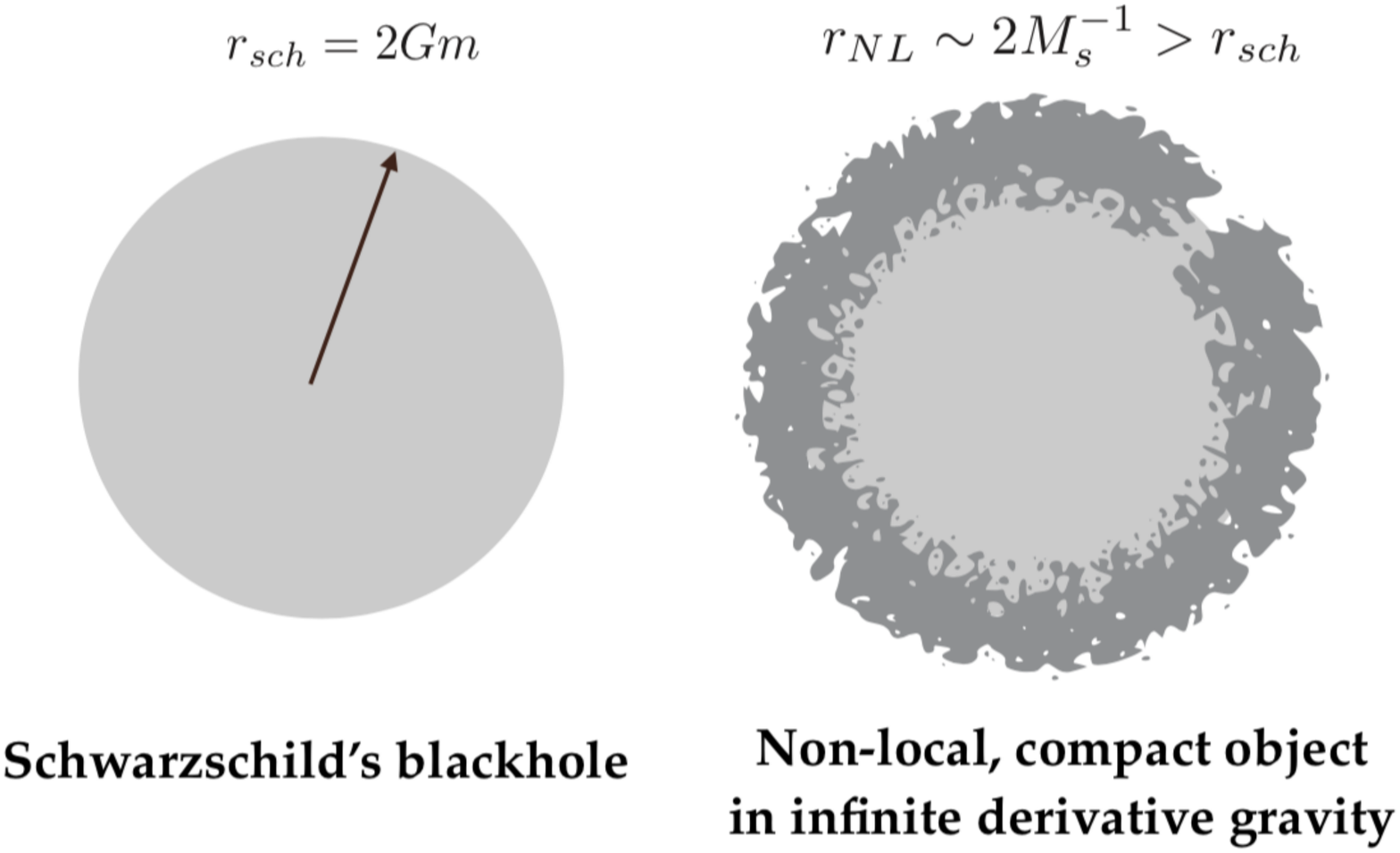}
 	\centering
 	\protect\caption{On the left side we have an illustration for a Schwarzschild blackhole in GR; while on the right side we have a non-local compact object which arises in infinite derivative gravity, where the non-local region engulfs the Schwarzschild radius, $r_{NL}\sim 2/M_s>r_{\rm sch}=2Gm.$ Inside the non-local region, neither clock no ruler makes sense, although an Euclidean description can be made compatible. Further note that in the GR the energy density would always blow up near the singularity, but in the infinite derivative case $\rho < M_p^4$ in the region of non-locality.}
 \end{figure}\label{fig1}

Another intriguing fact arises from counting the number of states confined in the non-local region. As we have argued in Ref.~\cite{Koshelev:2017bxd}, for astrophysical blackholes, the {\it effective} scale of non-locality moves towards the IR, $M_{s} \rightarrow M_s/\sqrt{N}\ll M_p$, where $N$ represents the number of gravitons confined inside the radius $1/M_s$; see also footnote \ref{shift}. Thus, as long as Eq.\eqref{eq:5} holds, the gravitational entropy goes as $S\sim S_q\sim N (M_p/M_s)^2$, and the number states ${\cal N}$ will be given by
\begin{equation}
{\cal N} \sim  e^{S_q}\sim e^{N(M_p/M_s)^2}\gg 1\,.
\end{equation}

Note that similar feature also arises in the fuzz-ball case where the number of available states are indeed {\it exponentially} large, and so is the Bekenstein entropy~\cite{Mathur:2005zp}.

$ $

{\it To summarize the results of this subsection: we have shown that non-locality is a necessary condition in order for the inequality in Eq.\eqref{eq:5} to be always satisfied, as one can introduce a scale $M_s$, which is not fixed, once infinite derivatives are taken into account. In particular, the inequality in Eq.\eqref{eq:5} is a non-perturbative statement.}


\subsection{Infinite derivatives acting on Dirac delta source}\label{smear-delta}

In this subsection, we will argue that the presence of infinite derivatives is crucial in order to avoid singular metric solutions. In particular, we will explain why the Schwarzschild metric with a singularity cannot be a solution for the full non-linear field equations in infinite derivative gravity with a boundary condition at $r=0$. This has been partly answered in Ref.\cite{Koshelev:2018hpt}, especially by studying the Weyl part of the equations of motion for Eq.(\ref{act}), and it was shown that the Scharzschild-like metric cannot be a vacuum solution of the field equations, as the contribution due to the Weyl part does not vanish. In Ref.\cite{Koshelev:2018rau}, it was also shown that {\it neither} the time-dependent Kasner-type metric (anisotropic collapsing Universe) is a solution for the infinite derivative gravity described by the action in Eq.\eqref{eq:1}. The full equations of motion for the above action has been derived in Ref.\cite{Biswas:2013cha}, which contains indeed {\it infinite covariant derivatives}.

Now, let us consider the field equations containing only the Ricci scalar and the Ricci tensor, by setting $\mathcal{F}_3(\Box_s)=0$ for the time being, and ask whether the Schwarzschild metric would be a solution in this case or not?  In such a case one can trivially notice that a Schwarzschild-like metric, i.e $g_{00}=-g^{11}=-1+2a/r,$ will solve the field equations in the vacuum at each order in $\Box_s$. However, we must note that in order to construct the Schwarzschild metric in GR, we also require a boundary condition by introducing a Dirac delta source at the origin - which implies a non-zero energy-momentum tensor at $r=0$, and this allows to fix $a=Gm.$ Thus, the right statement to make is that a Schwarzschild-like metric would solve the field equations with $\mathcal{F}_3(\Box_s)=0$, at each order in $\Box_s$, everywhere {\it except} at $r=0$. Such a boundary condition indeed plays a very crucial role in generating a singularity at $r=0$. 

We can now understand why the Schwarzschild metric cannot be a solution for the infinite derivative gravity even if $\mathcal{F}_3(\Box_s)=0$. It has been known that a finite number of spatial derivatives acting on a Dirac delta distribution {\it still} yields a point-source~\cite{Balasin:2006cg}. Would one repeat the same argument mathematically for any local theory of gravity, including quadratic curvature Stelle's gravity~\cite{Stelle}, one would obtain in the right hand side of the equations of motion, a  schematic form of $\delta(r)+\delta'(r)$, which is still a point-like source, but a bit less conventional one. Going further, more but finite number of derivatives will generate a finite series like
$$\delta(r)+\delta'(r)+\delta''(r)+\dots+\delta^{(N)}(r),$$ for some finite $N$. As long as N is finite we will still have a point-like source and the singular Schwarzschild metric is still a valid solution, even at $r=0$ with the boundary condition.

However, if we continue up to the infinite order in derivatives, then one would obtain a series with infinitely many terms.
By consulting the theory of distributions in this subtle case one can easily find out (see
books by Vladimirov~\cite{Vladimirov}, or by Gelfand and Shilov \cite{Gelfand} for example) that
a series of infinitely many terms with derivatives acting on the Dirac delta distribution corresponds
to a function with a non-point support. To see this explicitly, let us consider the following distribution on a real axis:
$$e^{\alpha\partial_x^2}\delta(x),$$
where $\alpha$ is a constant.
In order to understand this function, let us employ the Fourier transform, indeed this yields:
\begin{equation}
	e^{\alpha\partial_x^2}\delta(x)=\frac1{\sqrt{2\pi}}e^{\alpha\partial_x^2}\int dk e^{ikx}=\frac1{\sqrt{2\pi}}\int dk e^{-\alpha k^2} e^{ikx}
	=\frac1{\sqrt{2\alpha}}e^{-\frac{x^2}{4\alpha}}\,,
\end{equation}
which is manifestly a regular function with a non-point support. We have seen this argument before in Refs.~\cite{Siegel:2003vt,Biswas:2005qr,Biswas:2011ar} that an infinite tower of derivatives can smear the singular behavior of a Dirac delta mass distribution. Having in mind that in Einstein's GR we equate geometrical quantities to stress energy momentum tensor, we immediately conclude that any attempt to demand a singular metric potential like $\Phi(r)\sim 1/r$ or $1/r^{\alpha}$, where $\alpha>0$, i.e. a singular geometry, in the infinite derivative gravity will yield a regular matter. Or in other terms, a singular matter source would yield a regular gravitational potential~\cite{Tseytlin:1995uq,Siegel:2003vt,Biswas:2005qr,Biswas:2011ar}\footnote{For example the gravitational potential in Eq.\eqref{eq:8} is a solution of the modified Poisson equation $e^{-\nabla^2/M_s^{2}}\nabla^2\Phi(r)=4\pi Gm\delta^{(3)}(r)$, which can be also rewritten as $$\nabla^2\Phi(r)=4\pi Gm e^{\nabla^2/M_s^2 }\delta^{(3)}(r)=\frac{GmM_s^3}{2\sqrt{\pi}}e^{-\frac{M_s^2r^2}{4}}.$$ We can notice that the presence of infinite derivative has smeared out the Dirac delta source and produced a Gaussian-like distribution.}. 

This simple observation again corroborates the fact that the infinite derivative gravity can smoothen out the notion of space-time by imposing a fundamental length
$$r_{NL}\sim2/M_s,$$ such that one cannot probe physics classically, neither geometry nor matter at distances less than $r_{NL}$ or, equivalently, at energies greater  than $M_s$.

The usual notion of vacuum solution, that applies for the Schwarzschild metric in GR, does not apply in  the infinite derivative gravity, where as soon as we introduce a point-like source it smears the spacetime region by the scale of non-locality $r_{NL}\sim 2/M_s.$ In this respect, in the infinite derivative gravity a metric solution can be really called as a vacuum {\it if and only if} $T_{\mu\nu}=0$, everywhere in the space-time (including $r=0.$)  

$ $

{\it To summarize the results of this subsection: we have shown that the main feature of non-locality is given by the fact that infinite derivatives smear out a point-like delta source, implying that the Schwarzchild metric cannot be a solution for the infinite derivative gravity.}


\subsection{Linear perturbations: from IR to UV}\label{lin-sol}

In the linear regime for the above action in Eq.(\ref{eq:1}), we can determine the spherically symmetric static metric around the Minkowski background and analyze the structure of the gravitational potential. In Ref.\cite{Biswas:2011ar} it was shown that around Minkowski background the three form-factors can be constrained by requiring the absence of ghost-like degrees of freedom\footnote{In the original paper \cite{Biswas:2011ar} the ghost-free condition was formulated for the gravitational action with the Riemann tensor. In such a case the ghost-free condition would read $2F_1(\Box_s)+F_2(\Box_s)+2F_3(\Box_s)=0$, where the form-factors $F_i's$ have been defined in the footnote \ref{foot:1}.}$^{,}$\footnote{The propagator around Minkowski space-time, compatible with the condition in Eq.\eqref{eq:6}, reads \cite{Biswas:2011ar}
	\begin{equation}
	\Pi(k)=\frac{1}{a(k)}\left(\frac{\mathcal{P}^{2}}{k^2}-\frac{\mathcal{P}_s^{0}}{2k^2}\right),
	\end{equation}
	where $\Pi_{GR}(k)=\frac{\mathcal{P}^{2}}{k^2}-\frac{\mathcal{P}_s^{0}}{2k^2}$ is the GR graviton propagator, and $\mathcal{P}^2,\mathcal{P}^0_s$ are the so called spin-projector operators \cite{Biswas:2011ar,Biswas:2013kla,Buoninfante}. It is now clear that by demanding the form-factor in the last equation to be exponential of an entire function, $a(k)=e^{\gamma(k/M_s)}$, {\it no} extra poles would be introduced in the propagator other than the massless transverse-traceless graviton degree of freedom, so that tree-level unitarity will be preserved.}:
\begin{equation}
6\mathcal{F}_1(\Box_s)+3\mathcal{F}_2(\Box_s)+2\mathcal{F}_3(\Box_s)=0,\,\,\,\,\,\,\,\,\,\,\,a(\Box_s)=1+2\mathcal{F}_2(\Box_s)\Box_s+4\mathcal{F}_3(\Box_s)\Box_s=e^{\gamma(\Box_s)},\label{eq:6}
\end{equation}
in particular we consider the simplest choice $\gamma(\Box_s)=-\Box_s$; for other choices of  entire functions, different from $e^{-\Box_s}$, see Refs.\cite{Edholm:2016hbt}. By working in the weak-field regime, i.e. $2|\Phi(r)| < 1$, the linearized metric in the isotropic coordinates is given by \cite{Biswas:2011ar}
\begin{equation}
ds^2=-(1+2\Phi(r))dt^2+(1-2\Phi(r))[dr^2+r^2d\Omega^2],\label{eq:7}
\end{equation}
where the gravitational potential is defined as~\cite{Biswas:2011ar}
\begin{equation}
\Phi(r)=-\frac{Gm}{r}{\rm Erf}\left(\frac{M_s r}{2}\right)\,. \label{eq:8}
\end{equation}
In the IR regime, i.e. at large distances, the potential in Eq.\eqref{eq:8} recovers the Newtonian $1/r$-fall, while in the non-local regime, i.e. at short distances, the gravitational potential approaches a finite constant value:
\begin{equation}
r\ll 2/M_s \Longrightarrow \Phi(r)\sim \frac{GmM_s}{\sqrt{\pi}}=\frac{mM_s}{\sqrt{\pi}M_p^2}. \label{eq:9}
\end{equation}
The linear regime,  $2|\Phi(r)|<1$, implies 
\begin{equation}\label{lin}
mM_s<M_p^2\,,
\end{equation}
 which is nothing but the same inequality we have obtained in Eq.\eqref{eq:5}. Thus, we have shown that the inequality in Eq.\eqref{eq:5} holds {\it always} true in the linear regime. Indeed, if the entire astrophysical blackhole geometry becomes linear for $ r < r_{NL}\sim 2/M_{s} > r_{sch}$, such that $mM_{s} < M_p^2$, then there will be no horizon at all, and the metric potential throughout the geometry remains regular, for the entire range $0\leq r\leq \infty$.

In fact, for Eq.(\ref{eq:8})  all linearized curvature tensors have been computed and it was found that $\mathcal{R},\mathcal{R}_{\mu\nu}\neq 0$, meaning that the Schwarzschild vacuum solution of Einstein's GR is {\it not} a vacuum solution for the infinite derivative gravity; see Ref.\cite{Buoninfante:2018xiw}. The Ricci scalar, the Ricci tensor and the Riemann tensor tend to constant finite values at short-distances, i.e. for $r<2/M_s$. Moreover, all curvature invariants were shown to be non-singular, in particular the Kretshmann tensor turns out to be finite at $r=0$.  As for the Weyl tensor, it tends quadratically to zero in the non-local regime, implying that the space-time metric becomes {\it conformally-flat} in the region of non-locality; see Ref.\cite{Buoninfante:2018xiw}. 

Let us also point out that the metric in Eqs.(\ref{eq:7},\ref{eq:8}) is expressed in terms of the isotropic radial coordinate $r$. This metric can be rewritten in terms of the Schwarzschild coordinates by making the following transformation: 
\begin{equation}
R^2=(1-2\Phi(r))r^2\,,
\end{equation}
with $R$ being the radius in the Schwarzschild coordinates. Such a transformation can be easily inverted in the linear regime, indeed one can use the fact that up to first order in perturbations, we have 
\begin{equation}
\Phi(r)\simeq \Phi(R)\,, ~~~~{\rm and}~~~~~~~~~~dr\simeq[1+\Phi(R)+\Phi'(R)R]dR,
\end{equation}
 such that the metric in the new coordinates assumes the following form:
\begin{equation}
ds^2=-(1-2\Phi(R))dt^2+(1+2\Psi(R))dR^2+R^2d\Omega^2,\label{eq:8.2}
\end{equation}
where we have {\it two} different metric potentials:
\begin{equation}
\Phi(R)=-\frac{Gm}{R}{\rm Erf}\left(\frac{M_sR}{2}\right),\,\,\,\,\,\,\,\,\,\,\,\,\Psi(R)=-\frac{Gm}{R}{\rm Erf}\left(\frac{M_sR}{2}\right)+\frac{GmM_s}{\sqrt{\pi}}e^{-\frac{M_s^2R^2}{4}}.\label{two-pot}
\end{equation}
We can now notice a crucial difference between the static metric in GR and in infinite derivative gravity by looking at the form of the metric in Eqs.(\ref{eq:8.2},\ref{two-pot}). In GR the $00$- and $11$-components are not independent, i.e. $g_{00}=-g^{11}$,
where there is only one metric potential $-Gm/R$~\footnote{Indeed, the linearized Schwarzschild metric in Schwarzschild coordinates is given by:
\begin{equation}
ds^2=-\left(1-\frac{2Gm}{R}\right)dt^2+\left(1+\frac{2Gm}{R}\right)dR^2+R^2d\Omega^2.
\end{equation}
}, while in infinite derivative gravity, as one can notice from Eq.\eqref{two-pot}, the two metric components are independent, $g_{00}\neq-g^{11}$.

$ $

{\it To summarize the results of this subsection: we have shown that in the linear regime, by demanding the ghost-free condition through the introduction of form factors made of exponential of entire functions, the metric solution has no horizon, is regular and becomes conformally-flat at the origin, with no curvature singularity, moreover the inequality in Eq.\eqref{eq:5} is always satisfied. This means that in infinite derivative gravity one would expect the gravitational potential in the Schwarzschild coordinates or in the isotropic coordinates to be always bounded by $2|\Phi|<1,$ in the entire spacetime.}

\section{Requirements for a non-singular metric solution}


\subsection{Classical arguments for a conformally-flat core}

It is important to note the key parameters which govern the solution for the full linear and nonlinear action in Eq.~(\ref{eq:1}), and to understand how the infinite derivative gravitational interaction should behave in the regime where $r< 2/M_s$.

\begin{itemize}

\item {\bf The gravitational constant $G=1/M_p^2$} : it remains the same, so neither linear nor non-linear solutions are sensitive to $M_p$, the only constraint we have is $M_p\geq M_s$.

\item {\bf The mass of the system $m$} : the value of $m$ can indeed change in the range $0<m<\infty$. The question remains what happens when we are away from the weak-field regime, i.e. $mM_s > M_p^2$. Indeed, as we have already discussed above, violating this inequality would also demand that we have $S_{EH}> S_q$ within the Schwarzschild radius. In this case, the equations of motion due to infinite derivatives do not play any role, and therefore, singularity is inevitable. As we have seen in the previous section, finite derivatives acting on a point source would match the boundary condition for a Schwarzschild-like solution very well.

\item {\bf The scale of non-locality $M_s$ }: It has been argued that $M_s$ is not a fixed quantity, but it gets modified as the number of quanta in an enclosed area, or a volume, increases, i.e. as $m$ increases $M_s$ decreases, such that $mM_{s} < M_p^2$ is always satisfied for any range of $ 0<  m < \infty$~\cite{Koshelev:2017bxd}.  This physical argument in essence is similar to what we have proposed in this paper, where $S_q$ always dominates over $S_{EH}$ inside the region $r_{NL}\sim 2/M_s> r_{sch}$, such that $2|\Phi(r)|< 1$, and the metric potential tends to be {\it constant} as $r\rightarrow 0$. Indeed this case always leads to the resolution of a singularity, and also the avoidance of an event horizon. The space-time metric becomes conformally-flat, with a Riemann tensor still non-vanishing, therefore the space-time does feel the gradient of the metric potential, but the force on any particle vanishes in the limit $r \rightarrow 0$.

\end{itemize}

In the non-local region, $r < 2/M_s$, we can study the full equations of motion~\cite{Biswas:2013cha}, but in full generality they are extremely challenging to solve, therefore it is wise to study them and build the metric solution by using solid and valid physical arguments. Let us consider a general spherically symmetric metric of the following form: 
\begin{equation}
ds^2=-A(r)dt^2+B^{-1}(r)dr^2+r^2d\Omega^2.\label{gen-met}
\end{equation}
In our previous study, we have ruled out metric solutions of the form $A(r)=B(r)=1+2\Phi(r)$,\footnote{In Ref.\cite{Koshelev:2018hpt} it was considered the case $A(r)=B(r)$ but, more generally, we can argue that the same result also holds for static spherically symmetric metrics with $A(r)\neq B(r),$ namely with the presence of two gravitational potentials $\Phi(r)$ and $\Psi(r)$ such that $A(r)=1+2\Phi(r)$ and $B(r)=1+2\Psi(r)$.} with $\Phi(r)\sim 1/r^{\alpha}$ and $\sim r^{\beta}$, for $\alpha, \beta >0$, as possible full vacuum solutions for the  infinite derivative gravity, where the energy momentum tensor is vanishing in the entire region of spacetime~\cite{Koshelev:2018hpt}, due to the presence of the Weyl contribution. Furthermore, as we have already argued above, even in the absence of the Weyl term, Schwarzschild-like metrics would not pass through the field equations as infinitely many derivatives would smear out the delta source at $r=0.$ This last point is crucial. In fact, we can imagine other solutions, with explicit $r$ dependence in the regime $r \ll 2/M_s$, with boundary condition at $r=0.$ However, any physical solution which is continuous in the entire domain and infinitely differentiable at each and every point, can be Taylor expanded in the vicinity of $r \ll 2/M_s$, except if the function has a singularity at any point in the space-time, i.e. at $r=0$. In the latter case, for any such singular function, the leading order term in the metric potentials ought to be of the form $\Phi(r)\sim 1/r^{\alpha}$ ($\alpha >0$), but, again, we can argue that such a singular metric potential will be regularized at $r=0$ due to non-local gravitational interaction, preventing singular short-distance behavior. Such a scenario perfectly agrees with our expectation from the linearized regime.

We will now wish to build a metric solution and find {\it conditions} on how the two components $A(r)$ and $B(r)$ have to look like in order for the metric to approach to a {\it constant} value in the non-local region, i.e. for $M_sr\rightarrow 0$, such that the Weyl tensor tends to zero and the other curvature invariants to finite constant values, therefore matching the expectations from the linear regime \cite{Buoninfante:2018xiw}. 

In full generality, we will show that having a regular metric is {\it not} a sufficient condition to avoid curvature singularities, but we will need further constraints on the form of the metric components $A(r)$ and $B(r)$. Indeed, having a non-local action with infinite derivatives is {\it not} a sufficient condition to have non-singular metric solutions, namely not every kind of infinite derivative form-factors, $\mathcal{F}_i(\Box_s)$, would be suitable to resolve the curvature singularity. We will notice that the ghost-free condition in Eq.\eqref{eq:6} plays a crucial role to select the {\it right} form-factors.


\subsection{Conformally-flat non-local region}\label{conformal-core}

In this subsection, we want to study what are the conditions that a regular metric has to satisfy in order to become conformally-flat with constant curvature invariants in the short-distance regime, $r\ll 2/M_s$.  From Eqs.(\ref{eq:8.2},\ref{two-pot}) of the previous section, we have learnt that, unlike in GR, in infinite derivative gravity one has $A(r)\neq B(r)$ and one has to deal with the more general case of {\it two} metric potentials $\Phi(r)$ and $\Psi(r)$. It means that the Birkhoff's theorem is violated in infinite derivative gravity in the region of non-locality. Thus, for our discussion we will consider the general case of two different metric components in Eq.\eqref{gen-met}:
\begin{equation}
A(r)=1+2\Phi(r),\,\,\,\,\,\,\,\,\,\,\,\,\,\,\,B(r)=1+2\Psi(r),\label{two-pot2}
\end{equation}
where $r$ now stands for the radial coordinate in the Schwarzschild coordinates. As we have already demanded, $A(r)$ and $B(r)$ have to be non-singular within the non-local region, i.e. $r < 2/M_s$, therefore the Taylor expansion at $r=0$ yields:
\begin{equation}
\Phi(r)=\Phi_0+\Phi_1 r+\frac{1}{2}\Phi_2 r^2+\frac{1}{6}\Phi_3 r^3+\cdots,\,\,\,\,\,\,\,\,\,\,\,\Psi(r)=\Psi_0+\Psi_1 r+\frac{1}{2}\Psi_2 r^2+\frac{1}{6}\Psi_3 r^3+\cdots, \label{expans}
\end{equation}
such that for $r\rightarrow0$ we obtain a regular behavior, $\Phi\rightarrow \Phi_0$ and $\Psi\rightarrow\Psi_0$. All coefficients in the expansions in Eq.\eqref{expans} are ought to be functions of $G$, $m$ and $M_s$, in particular, by dimensional analysis, we would expect:
\begin{equation}
\Phi_n\sim GmM_s^{n+1},\,\,\,\,\,\,\,\,\,\,\,\,\Psi_n\sim GmM_s^{n+1}. \label{expans.2}
\end{equation}
A non-singular space-time metric at the origin is {\it not} a sufficient condition in order to avoid a curvature singularity; in fact, by looking at the curvature invariants in the regime, $M_sr\ll 1$, we note that for generic values of the coefficients in Eq.\eqref{expans}, we would still have curvature singularities. Indeed, we can compute the curvature invariants for the metric defined in terms of the potentials in Eq.\eqref{expans}. The Ricci scalar in the non-local region, i.e. $M_sr\ll 1$, has the following structure:
\begin{equation}
\mathcal{R}\sim -\frac{4\Psi_0}{r^2}+\frac{f_1(\Phi_i,\Psi_i)}{r}+f_0(\Phi_i,\Psi_i)+\mathcal{O}(r);
\end{equation}
the Ricci tensor squared:
\begin{equation}
\mathcal{R}_{\mu\nu}\mathcal{R}^{\mu\nu}\sim \frac{8\Psi_0^2}{r^4}+\frac{g_3(\Phi_i,\Psi_i)}{r^3}+\frac{g_2(\Phi_i,\Psi_i)}{r^2}+\frac{g_1(\Phi_i,\Psi_i)}{r}+g_0(\Phi_i,\Psi_i)+\mathcal{O}(r);
\end{equation}
the Kretschmann invariant:
\begin{equation}
\mathcal{K}\equiv\mathcal{R}_{\mu\nu\rho\sigma}\mathcal{R}^{\mu\nu\rho\sigma}\sim \frac{64\Psi_0^2}{3r^4}+\frac{h_3(\Phi_i,\Psi_i)}{r^3}+\frac{h_2(\Phi_i,\Psi_i)}{r^2}+\frac{h_1(\Phi_i,\Psi_i)}{r}+h_0(\Phi_i,\Psi_i)+\mathcal{O}(r);
\end{equation}
and the Weyl squared:
\begin{equation}
\mathcal{C}_{\mu\nu\rho\sigma}\mathcal{C}^{\mu\nu\rho\sigma}\sim \frac{16\Psi_0^2}{3r^4}+\frac{c_3(\Phi_i,\Psi_i)}{r^3}+\frac{c_2(\Phi_i,\Psi_i)}{r^2}+\frac{c_1(\Phi_i,\Psi_i)}{r}+c_0(\Phi_i,\Psi_i)+\mathcal{O}(r),
\end{equation}
where the functions $f_n,g_n,h_n,c_n$ depend only on the coefficients $\Phi_i,\Psi_i$, with $i=0,\dots,4.$ Note that the curvature invariants can blow up at $r=0$ if we do not make further assumptions on the form of the coefficients in the metric potentials in Eq.\eqref{expans}. 

Given the above analysis, we can now deduce the conditions under which the metric inside the non-local region will be singularity free.

\begin{enumerate}

\item {\bf $\Psi_0=0$}: The first requirement for non-singular curvature invariants is that $\Psi_0=0$, which is also in agreement with the linearized metric in Eqs.(\ref{eq:8.2},\ref{two-pot}), as we note by making a direct comparison with the linearized form of the general metric in Eqs.(\ref{gen-met},\ref{two-pot2}), i.e. in the regime $2|\Phi(r)|,2|\Psi(r)|\ll1.$

\item {\bf $\Phi_1=0\,,~\Psi_1=0$}: Moreover, one can also check that all the above curvature invariants would become singularity-free by making the additional requirements $\Phi_1=0$ and $\Psi_1=0$; which are also in agreement with the linear regime, as we can see by expanding the two metric potentials in Eqs.(\ref{eq:8},\ref{two-pot}) around the origin. Therefore, we have:
\begin{equation}
\Phi_1=\Psi_0=\Psi_1=0\,\,\,\Longrightarrow\,\,\,f_n=g_n=h_n=c_n=0,\,\,\,{\rm for}\,\,\,n=1,2,3.\label{requir}
\end{equation}
\item As for the functions with $n=0$, it so happens that if $\Phi_1=\Psi_0=\Psi_1=0$, then $f_0,g_0,h_0$ are in general non-vanishing constants, while $c_0$ {\it vanishes}. This means that by assuming the metric in Eq.\eqref{gen-met} with the two regular metric potentials in Eq.\eqref{expans}, the requirement in Eq.\eqref{requir} turns out to be a {\it necessary and sufficient condition} in order to have a {\it conformally-flat regular metric with constant curvature invariants} in the limit $r\rightarrow0.$

In particular, in the limit $r\rightarrow0$ the Ricci scalar, the Ricci tensor squared and the Kretschmann invariant tend to finite constant values, while the Weyl tensor squared tends to zero\footnote{It can be explicitly checked that, given the condition in Eq.\eqref{requir}, all components of the Weyl tensor go to zero for $r\rightarrow 0.$}:
\begin{equation}
\mathcal{R}\sim f_0,\,\,\,\,\,\,\,\,\mathcal{R}_{\mu\nu}\mathcal{R}^{\mu\nu}\sim g_0,\,\,\,\,\,\,\,\,\mathcal{K}\sim h_0,\,\,\,\,\,\,\,\mathcal{C}_{\mu\nu\rho\sigma}\mathcal{C}^{\mu\nu\rho\sigma}\sim 0. \label{non-localcurv}
\end{equation}
\end{enumerate}
From the above analysis, it is very clear that for the regular metric in Eqs.(\ref{two-pot2},\ref{expans}), with $\Phi\neq\Psi,$ conformal-flatness and avoidance of curvature singularities imply each other. Indeed, if the metric is not conformally-flat for $r\rightarrow 0$, then the metric will show curvature singularities; while if the metric has curvature singularities at $r=0$, then the metric will not be conformally-flat~\footnote{Note that such a result is particularly based on the two following hypothesis;  (1) $g_{00}\neq -g_{11}$, and (2) $\Phi\neq \Psi$. The condition (1) is really fundamental in order to recover GR in the IR regime. However, for curiosity one can ask what happens if it were violated. In fact, if $g_{00}=-g_{11}$, we can have metrics which are conformally-flat for $r\rightarrow0$, but still the corresponding curvature invariants blow up at $r=0.$ As an example, we can consider the following metric with $$g_{00}=-g_{11}=-\left(\frac{r}{r+a}\right)^2,$$ and we can check that it becomes conformally-flat for $r\rightarrow0$, but the corresponding Kretschmann invariant blows up. Moreover, if (1) holds but (2) is violated, we must have $\Psi=\Phi$ which also means $g_{00}=-g^{11},$ and in such a case only one implication holds: non-singular curvature invariants $\Longrightarrow$ conformal-flatness. Indeed, in this case, one can check that the condition $\Phi_0=0$ ensures conformal-flatness, but in order to have non-singular curvature invariants, we would need to make the additional requirement, that is $\Phi_1=0$. As an example we can consider a metric with $\Phi=a r,$ where $a$ is a constant, which turns out to be conformally-flat for $r\rightarrow0$, but other curvature invariants, like the Kretschmann, blow up at $r=0$, this can be easily checked. However, we now know that in infinite derivative gravity one has to deal with two different metric potentials $\Phi\neq\Psi$ within the scale of non-locality, and in such a case, for a regular metric, conformal-flatness and non-singular curvature invariants come hand in hand.}.

From the above analysis, it is also clear that having a regular metric is not a sufficient condition to avoid curvature singularities, but we would need to make the additional requirements $\Psi_0=\Psi_1=\Phi_1=0$, so that we would require $\Phi(r)\neq \Psi(r) $ for $ r < 2/M_s$ and $\Phi(r)\sim \Psi (r) $ for $r\gg 2/M_s$. We indeed confirm that these requirements are {\it exactly} in agreement with the linearized metric of infinite derivative gravity in Eqs.(\ref{eq:8.2},\ref{two-pot}); the leading order coefficients for the linearized metric are given by:
\begin{equation}
\Phi_0=-\frac{GmM_s}{\sqrt{\pi}},\,\,\,\,\,\,\,\,\Phi_1=0,\,\,\,\,\,\,\,\,\Phi_2=\frac{GmM_s^3}{6\sqrt{\pi}},\,\,\,\,\,\,\,\,\Psi_0=\Psi_1=0,\,\,\,\,\,\,\,\,\Psi_2=-\frac{GmM_s^3}{3\sqrt{\pi}}.
\end{equation}
We can now understand that the {\it ghost-free condition} has played a crucial role in selecting the right metric potentials $\Phi$ and $\Psi$. In fact, in general one can choose form-factors $\mathcal{F}_i$ in the action in Eq.\eqref{eq:1} such that they can yield non-singular metric solutions, despite the presence of non-locality through infinitely many derivatives. The choice of {\it exponential of entire functions}, which is required  to avoid ghost-like degrees of freedom, also constrain the metric potentials, such that the conditions in Eq.\eqref{requir} are satisfied. In the previous section, we have considered the case $\gamma(\Box_s)=-\Box_s$, but the same scenario is kept for $\gamma(\Box_s)$ being a generic polynomial of $\Box_s$. 

$ $

{\it Indeed, in infinite derivative gravity astrophysical blackholes can be free from singularity and event horizon provided within the scale of non-locality the metric solution yields $\mathcal{R}\sim f_0,\,\mathcal{R}_{\mu\nu}\mathcal{R}^{\mu\nu}\sim g_0,\,\mathcal{K}\sim h_0,\,\mathcal{C}_{\mu\nu\rho\sigma}\mathcal{C}^{\mu\nu\rho\sigma}\sim 0$, in the limit $r\rightarrow 0$.}

\section{Conclusion}

Typical Schwarzschild metric within Einstein-Hilbert action relies on a non-trivial vacuum solution, which yields $1/r$ metric potential, and the boundary condition at $r=0$, where there exists a central mass with the Dirac delta distribution. Except $r=0$, we have a vacuum solution. This very fact can be modified manifestly in covariant formulation of {\it ghost-free} infinite derivative theory of gravity, which is the key concept that ameliorates the curvature singularity and the avoidance of an event horizon for any blackhole mass. The non-perturbative statement due to 
Eq.\eqref{eq:5} affirms that the metric potential in the entire region of non-locality with $r\sim 2/M_s> r_{sch}$, for any  astrophysical source mass $m$, is such that the condition 
$mM_s < M_p^2$ is always satisfied. Moreover, the metric potentials $\Phi$ and $\Psi$ are such that the spacetime metric becomes conformally-flat and free from curvature singularity in the region $r\ll 2/M_s$.

As we have already mentioned above, the ghost-free condition ensures that the requirements in Eq.\eqref{requir} hold, which also imply that for two different regular metric potentials, $\Phi$ and $\Psi$, conformal-flatness is a necessary and sufficient condition for non-singular curvature invariants. It means that in a ghost-free, infinite derivative theory of gravity we cannot have a spacetime metric which is conformally-flat but suffers from curvature singularity, or vice versa; indeed, the two requirements imply each other. 

We have also argued from a purely mathematical point of view that the Schwarzschild metric, for example, will fail to solve our full equations of motion, containing infinite derivatives due to boundary condition. This is due to the fact that any point-like source at $r=0$ is smeared out on a region whose size is given by the scale of non-locality. Our study sheds very important light on infinite derivative, ghost-free, theories of gravity as proposed in Ref.\cite{Biswas:2011ar}, with form factors determined by having no zeroes,  or no extra poles in the propagator, or no extra dynamical degrees of freedom other than the massless graviton, for maintaining the unitarity constraint on such an action. This quadratic curvature action has a potential to modify the way we view astrophysical blackholes, in particular, one of the key modifications will be that the astrophysical blackholes are no longer classical objects. Astrophysical blackholes remain compact with radius roughly given by the scale of non-locality, i.e.
$r_{NL}\sim 2/M_s> r_{sch}$. At distances larger than this radius the metric follows exactly that of GR, in the static and spherically symmetric coordinates. For all practical purposes, a distant observer will view it as a Schwarzschild metric, but as we approach closer to the scale of non-locality, we would realize that the metric approaches to be conformally-flat and the spacetime metric becomes non-local and free from any curvature singularity.

Indeed, this opens a new avenue for future works, such as studying quasi-normal modes, grey body factor, and put constraints on such non-singular compact objects from observational perspectives, such as gravitational wave observatories like LIGO and VIRGO in the near future.


\acknowledgments 
The authors would like to thank Valeri Frolov and Alexei Starobinsky for various discussions. AM would also like to thank Samir Mathur for many discussions on fuzz-ball scenario.
AK and JM are supported by the grant UID/MAT/00212/2013 and COST Action CA15117 (CANTATA). AK is supported by FCT Portugal investigator project IF/01607/2015 and FCT Portugal fellowship SFRH/BPD/105212/2014. 


\appendix

\section{Conformally-flat, non-linear vacuum solutions in the non-local regime}

In the main text we have mainly focused on the metric solution in infinite derivative gravity which generalizes the Schwarzschild-like metric to the case of non-local gravitational interaction. In this respect, we have considered a {\it non-vacuum} solution for the infinite derivative gravity field equations where the stress-energy tensor is non-vanishing inside the region of non-locality $r_{NL}\sim 1/M_s,$ due to the smearing of the Dirac delta source.

However, it can be also interesting to investigate the full field equations for the action in Eq.\eqref{eq:1} and seek {\it purely} vacuum static solutions in the region of non-locality, where the Einstein-Hilbert contribution can be neglected, $S_q\gg S_{EH},$ without imposing any boundary condition at $r=0.$ 
In particular, we wish to seek vacuum solutions which are conformally-flat in the region of non-locality:
\begin{equation} 
ds^2=F^2(r)[-dt^2+dr^2+r^2d\Omega^2],~~~~~~~{\rm for}~~r\ll1/M_s\,.  \label{eq:14}
\end{equation}
Of course, since we are not putting any Dirac delta source, such a scenario will not necessarily match the expectations of the linear regime, where one approaches to both constant curvature invariants and conformal-flatness. In order to make this study, we need to focus on the non-local part of the field equations corresponding to the action in Eq.\eqref{eq:1} and impose the ansatz of conformally-flat solution, so that the Weyl contribution vanishes in the non-local region, i.e. $C^{\mu\nu\rho\sigma}\rightarrow 0$ in the limit $r\rightarrow 0$. 

The complete field equations for the action in Eq.~\eqref{eq:1} were derived in Ref.\cite{Biswas:2013cha}, but we do not need their full expression for what follows. In fact, by focusing in the non-local regime, $S_q\gg S_{EH}$, the Einstein-Hilbert term can be neglected and by assuming conformally-flat metric solutions as ansatz, the Weyl contribution can be set to zero, $\mathcal{O}(\mathcal{C})$=0. The {\it only} relevant part of the field equations needed to seek conformally-flat vacuum solutions in the region $r<2/M_s$ are given by\footnote{Note that we are keeping the Einstein-Hilbert term in our action, without that just the quadratic action alone will have massive spin-2 ghost. Also, keep in mind that in the IR, we must recover the standard predictions of GR, i.e. the Schwarzschild vacuum solution outside $r>2/M_s$. Here we are interested in studying the limit where $S_q>S_{EH}$, or $r<2/M_s$.}
\begin{align}
P^{\alpha\beta}\approx& \frac{\alpha_c}{{ 8\pi G}} \biggl( 4G^{\alpha\beta}{\cal
	F}_{1}(\Box_s)\mathcal{R}+g^{\alpha\beta}\mathcal{R}{\cal
	F}_1(\Box_s)\mathcal{R}-4\left(\nabla^{\alpha}\nabla^{\beta}-g^{\alpha\beta}
\Box\right){\cal F}_{1}(\Box_s)\mathcal{R}
\nonumber\\&
-2\Omega_{1}^{\alpha\beta}+g^{\alpha\beta}(\Omega_{1\sigma}^{\;\sigma}+\bar{
	\Omega}_{1}) +4\mathcal{R}_{\mu}^{\alpha}{\cal F}_2(\Box_s)\mathcal{R}^{\mu\beta}
\nonumber\\&
-g^{\alpha\beta}\mathcal{R}_{\nu}^{\mu}{\cal
	F}_{2}(\Box_s)\mathcal{R}_{\mu}^{\nu}-4\nabla_{\mu}\nabla^{\beta}({\cal
	F}_{2}(\Box_s)\mathcal{R}^{\mu\alpha})
+2\Box({\cal
	F}_{2}(\Box_s)\mathcal{R}^{\alpha\beta})
\nonumber\\&
+2g^{\alpha\beta}\nabla_{\mu}\nabla_{
	\nu}({\cal F}_{2}(\Box_s)\mathcal{R}^{\mu\nu})
-2\Omega_{2}^{\alpha\beta}+g^{\alpha\beta}(\Omega_{2\sigma}^{\;\sigma}+\bar{
	\Omega}_{2}) -4\Delta_{2}^{\alpha\beta}
\biggr)
\nonumber\\
= & T^{\al\bt}=0\,,
\label{EOM}
\end{align}
where the symmetric tensors are given by~\cite{Biswas:2013cha}:
\begin{align}\label{details}
\Omega_{1}^{\alpha\beta}= & \sum_{n=1}^{\infty}f_{1_{n}}\sum_{l=0}^{n-1}\nabla^{
	\alpha}\mathcal{R}^{(l)}\nabla^{\beta}\mathcal{R}^{(n-l-1)},\quad\bar{\Omega}_{1}=\sum_{n=1}^{\infty
}f_{1_{n}}\sum_{l=0}^{n-1}\mathcal{R}^{(l)}R^{(n-l)},
\\
\Omega_{2}^{\alpha\beta}= & \sum_{n=1}^{\infty}f_{2_{n}}\sum_{l=0}^{n-1}\mathcal{R}_{\nu}^{
	\mu;\alpha(l)}\mathcal{R}_{\mu}^{\nu;\beta(n-l-1)},\quad\bar{\Omega}_{2}=\sum_{n=1}^{
	\infty}f_{2_{n}}\sum_{l=0}^{n-1}\mathcal{R}_{\nu}^{\mu(l)}\mathcal{R}_{\mu}^{\nu(n-l)}\,,
\\
\Delta_{2}^{\alpha\beta}= & \sum_{n=1}^{\infty}f_{2_{n}}\sum_{l=0}^{n-1}
[\mathcal{R}_{
	\sigma}^{\nu(l)}\mathcal{R}^{(\beta\sigma;\alpha)(n-l-1)}-\mathcal{R}_{\;\sigma}^{\nu;\alpha(l)
}\mathcal{R}^{
	\beta\sigma(n-l-1)}]_{;\nu}\,.\label{details-1}
\end{align}
We adopt the notation $\mathcal{R}^{(l)}\equiv\Box^l\mathcal{R}$ for the curvature tensors and their covariant derivatives.

The trace equation in the non-local regime and for conformally-flat metric ansatz can be written as~\cite{Biswas:2013cha}:
\begin{align}
P \approx & \frac{\alpha_c}{{8\pi G}} \biggl( 12\Box{\cal F}_{1}(\Box_s)\mathcal{R}+2\Box({\cal
	F}_{2}(\Box_s)\mathcal{R})+4\nabla_{\mu}\nabla_{\nu}({\cal
	F}_{2}(\Box_s)\mathcal{R}^{\mu\nu})
\nonumber\\ &
+2(\Omega_{1\sigma}^{\;\sigma}+2\bar{\Omega}_{1})
+2(\Omega_{2\sigma}^{\;\sigma}+2\bar{\Omega}_{2})
-4\Delta_{2\sigma}^{\;\sigma} \biggr)
=  T\equiv
g_{\al\bt}T^{\al\bt}=0\,.
\label{trace}
\end{align}
The next step will be to substitute the ansatz Eq.\eqref{eq:14} into Eq.\eqref{EOM} in order to study the field equations for the unknown conformal factor $F(r)$, and see how it would look like at each {\it order} of the series expansion of the form-factors $\mathcal{F}_1(\Box_s)$ and $\mathcal{F}_2(\Box_s)$. An explicit calculation of the component $P^{00}$ is carried at zero $\Box_s$ and at one $\Box_s$. At just one $\Box_s$, we can already see that the expression becomes very complicated, involving derivatives of the conformal factor $F^{(n)}(r)\equiv d^{(n)}F(r)/dr^n$, originating a highly nonlinear differential equation. At the zero order in the expansion of $\mathcal{F}_1(\Box_s)$ and $\mathcal{F}_2(\Box_s)$, the $00$-component of the field equations in Eq.\eqref{EOM} reads:
\begin{align}
P^{00}_{(0)}=- & \frac{\alpha_{c}\left(3f_{10}+f_{20}\right)}{2\pi G\:r^{4}F(r)^{7}}
\biggl[32r^{3}F'(r)^{3}-40r^{3}F(r)F'(r)F''(r)\nonumber \\
& +8r^{3}F(r)^{2}F^{(3)}(r)-5r^{4}F(r)F''(r)^{2}+16r^{4}F'(r)^{2}F''(r)\\
& +2r^{4}F(r)^{2}F^{(4)}(r)-10r^{4}F(r)F'(r)F^{(3)}(r)\biggr]=0\nonumber
\end{align}
while at the first order, or in other words at one $\Box_s$,  
\begin{equation*}
\begin{split}P^{00}_{(1)}=- & \frac{3\alpha_{c}f_{11}}{2\pi G\:r^{6}F(r)^{11}M_{s}^{2}}
\biggl[576r^{4}F'(r)^{4}\left(r^{2}F''(r)+2rF'(r)\right)\\
& -r^{2}F(r)F'(r)^{2}\biggl(645r^{4}F''(r)^{2}+460r^{2}F'(r)^{2}\\
& +4rF'(r)\left(109r^{3}F^{(3)}(r)+707r^{2}F''(r)\right)\biggr)\\
& +2F(r)^{2}\biggl(33r^{6}F''(r)^{3}+3r^{3}F'(r)F''(r)\left(61r^{2}F^{(3)}(r)+172r^{2}F''(r)\right)\\
& -68r^{3}F'(r)^{3}+2r^{2}F'(r)^{2}\left(37r^{4}F^{(4)}(r)+219r^{3}F^{(3)}(r)+97r^{2}F''(r)\right)\biggr)\\
& -F(r)^{3}\Biggl(28r^{2}F'(r)^{2}+\biggl(23r^{6}F^{(3)}(r)^{2}+28r^{4}F''(r)^{2}\\
& +20\left(2r^{4}F^{(4)}(r)+11r^{3}F^{(3)}(r)\right)r^{2}F''(r)\biggr)\\
& +2rF'(r)\left(\left(13r^{5}F^{(5)}(r)+76r^{4}F^{(4)}(r)+14r^{3}F^{(3)}(r)\right)-28r^{2}F''(r)\right)\Biggr)\\
& +2F(r)^{4}\left(r^{6}F^{(6)}(r)+6r^{5}F^{(5)}(r)\right)\biggr]
\end{split}
\end{equation*}
\begin{equation}
\begin{split}
& -\frac{\alpha_{c}f_{21}}{2\pi G\:r^{6}F(r)^{12}M_{s}^{2}}\biggl[24r^{4}F(r)F'(r)^{4}\left(53r^{2}F''(r)+64rF'(r)\right)-324r^{6}F'(r)^{6}\\
& -r^{2}F(r)^{2}F'(r)^{2}\biggl(1025r^{4}F''(r)^{2}+446r^{2}F'(r)^{2}\\
& +2rF'(r)\left(283r^{3}F^{(3)}(r)+1702r^{2}F''(r)\right)\biggr)\\
& +2F(r)^{3}\biggl(47r^{6}F''(r)^{3}+4r^{3}F'(r)F''(r)\left(59r^{3}F^{(3)}(r)+153r^{2}F''(r)\right)\\
& -74r^{3}F'(r)^{3}+r^{2}F'(r)^{2}\left(79r^{4}F^{(4)}(r)+470r^{3}F^{(3)}(r)+180r^{2}F''(r)\right)\biggr)\\
& -F(r)^{4}\Biggl(30r^{2}F'(r)^{2}+\biggl(29r^{6}F^{(3)}(r)^{2}+30r^{4}F''(r)^{2}\\
& +2r^{2}F''(r)\left(23r^{4}F^{(4)}(r)+126r^{3}F^{(3)}(r)\right)\biggr)+2rF'(r)\biggl(-30r^{2}F''(r)\\
& +\left(13r^{5}F^{(5)}(r)+76r^{4}F^{(4)}(r)+10r^{3}F^{(3)}(r)\right)\biggr)\Biggr)\\
& +2F(r)^{5}\left(r^{6}F^{(6)}(r)+6r^{5}F^{(5)}(r)\right)\biggr] =0
\end{split}
\end{equation}

Of course, a trivial solution inside the non-local region will be given by a constant conformal factor\footnote{This is evident due to the fact a constant metric is always a solution provided there is no cosmological constant term.} 
\begin{equation}\label{favsol}
F(r)\sim {\rm const.}~~~{\rm for}~~r \ll 2/M_s\,.
\end{equation}
Let us now investigate, if we could seek other non-trivial non-linear vacuum solutions of the above field equations~\footnote{None of our arguments would modify if we were to study pure Euclidean metric in the above, or later stages of this paper.}.

\subsection{Solution of type: $F(r)=\dfrac{2}{M_s r}$}
\label{part sol}

Let us now consider the following conformally-flat metric as an ansatz:
\begin{equation}\label{ansol}
ds^{2}=\left(\frac{2}{M_{s}r}\right)^{n}\left(-dt^{2}+dr^{2}+r^2d\Omega^{2}\right)\,,
\end{equation}
and study the consequences for the equations of motion; let us first compute the corresponding curvature invariants.

The Ricci scalar is given by
\begin{equation}
\mathcal{R}=-(n-2)\:\frac{3\cdot2^{-n-1}n\left(M_{s}r\right)^{n}}{r^{2}}\,,
\end{equation}
which vanishes for $n=2$; while the Ricci tensor squared reads
\begin{equation}
\mathcal{R}^{\mu\nu}\mathcal{R}_{\mu\nu}=\frac{2^{-2(n+1)}n^{2}\left(n\left(3n-14\right)+20\right)\left(M_{s}r\right){}^{2n}}{r^{4}}\,,
\end{equation}
which becomes $\dfrac{M_{s}^{2}}{4}$ for $n=2$. The Kretschmann tensor is given by:
\begin{equation}
\mathcal{K}\equiv \mathcal{R}^{\mu\nu\rho\sigma}\mathcal{R}_{\mu\nu\rho\sigma}=\frac{2^{-2(n+1)}n^{2}\left(n\left(3n-16\right)+28\right)\left(M_{s}r\right){}^{2n}}{r^{4}}\label{kret}
\end{equation}
which becomes $\dfrac{M_{s}^{2}}{2}$ for $n=2$, being non-singular.

We have been able to explicitly evaluate the field equations for the metric ansatz in Eq. \eqref{ansol} up to second order in boxes. The $00$-component of the full field equations at zero order in box yields:
\begin{equation}
P_{\left(0\right)}^{00}=-(n-2)\:\dfrac{\alpha_{c}\left(M_{s}r\right){}^{2n}}{\pi Gr^{4}}\left[2n\left(3n^{2}-18n+16\right)\left(3f_{10}+f_{20}\right)\right]\,,
\end{equation}
at one box order, yields:
\begin{equation}
\begin{aligned}P_{\left(1\right)}^{00}= & (n-2)\:\dfrac{\alpha_{c}\left(M_{s}r\right){}^{3n}}{\pi Gr^{6}M_{s}^{2}}\Biggl[8^{-n-2}n\biggl(6\left(3n^{4}+2n^{3}-96n^{2}+256n-192\right)f_{11}+\\
& +\left(9n^{4}+6n^{3}-236n^{2}+560n-384\right)f_{21}\biggr)\Biggr]\,,
\end{aligned}
\end{equation}
and at two boxes order, yields:
\begin{equation}
\begin{aligned}P_{\left(2\right)}^{00}= & (n-2)^{2} \: \dfrac{\alpha_{c}\left(M_{s}r\right){}^{4n}}{\pi Gr^{8}M_{s}^{4}}\Biggl[2^{-4n-5}n\biggl(-3\left(100n^{4}-957n^{3}+3356n^{2}-5124n+2880\right)f_{12}+\\
& \left(25n^{5}-296n^{4}+1713n^{3}-4770n^{2}+6044n-2880\right)f_{22}\biggr)\Biggr]\,.
\end{aligned}
\end{equation}
We can immediately see that for $n=2$, all of them vanish, i.e., $P_{\left(0\right)}^{00},~P_{\left(1\right)}^{00},~P_{\left(2\right)}^{00}=0$. Would this solution survive if we were to consider the expansions of the form-factors $\mathcal{F}_1$ and $\mathcal{F}_2$ at higher orders in $\Box_s$? We certainly have indications that it will, since the relevant pieces to compute at higher orders in Eq. (\ref{EOM}) have a common factor:
\begin{equation}
\Box_s R^{\mu \nu} \sim (n-2)..., \quad \nabla_{\alpha} R^{\mu \nu} \sim (n-2)...
\end{equation}
which is $(n-2)$. 

It would be important to understand this special non-trivial solution which yields conformally-flat metric solution in the non-local regime, but with vanishing scalar curvature, and which turns out to be non-singular as the Kretschmann invariant is just a constant; see Eq.\eqref{kret} with $n=2.$

\subsection{Other variants of non-local action}

We now wish to consider simplified version of the gravitational action in Eq.\eqref{eq:1}, in which we set one of the form-factors to be zero. Note that the present study focuses on conformally-flat metrics, and as such the Weyl squared term in the action is irrelevant for the field equations. The contribution with the form factor $\mathcal{F}_3(\Box_s)$ does not play any role for this analysis; it is however responsible for making the theory unitary at a quantum level. Other terms do show their importance already at the background level. Below we will consider two particular cases in which we set one of the two form-factors $\mathcal{F}_1(\Box_s)$, $\mathcal{F}_2(\Box_s)$ to zero, and analyze the consequences.

\begin{itemize}

	\item If we set $\mathcal{F}_2(\Box_s)=0$, we would deal with the following gravitational action\footnote{It is worth mentioning that such an action was considered in particular in Ref.\cite{Koshelev:2017tvv}, where the authors have shown that it is the most general quadratic action for studying linear perturbations around maximally symmetric backgrounds up to second order in the perturbations; i.e. the $\mathcal{F}_2(\Box_s)$ term would be redundant in such a case.}
	\begin{equation}
	S= \displaystyle \frac{1}{16\pi G}\int d^4x\sqrt{-g}\left[\mathcal{R}+\alpha_c\left(\mathcal{R}\mathcal{F}_1(\Box_s)\mathcal{R}+\mathcal{C}_{\mu\nu\rho\sigma}\mathcal{F}_{3}(\Box_s)\mathcal{C}^{\mu\nu\rho\sigma}\right)\right]. \label{choice2}
	\end{equation}
	We can look for a conformally-flat solution in the non-local regime, but now for the simpler action in Eq. \eqref{choice2}. The corresponding vacuum field equations in the region $r<2/M_s$ reads 
	\begin{equation}
	\begin{array}{rl}
	P^{\alpha \beta}\approx& \displaystyle \frac{\alpha_c}{8\pi G}\left[4G^{\alpha \beta}\mathcal{F}_1(\Box_s)\mathcal{R}+g^{\alpha \beta}\mathcal{R}\mathcal{F}_1(\Box_s)\mathcal{R}-4(\nabla^{\alpha}\partial^{\beta}-g^{\alpha \beta}\Box)\mathcal{F}_1(\Box_s)\mathcal{R}\right.\\
	& \displaystyle \,\,\,\,\,\,\,\,\,\,\,\,\,\,\,\,\,\,\,\,\,\,\,\,\,\,\,\,\,\,\,\,\,\,\,\,\,\,\left. -2\Omega_1^{\alpha \beta}+g^{\alpha\beta}(\Omega_{1\sigma}^{\sigma}+\bar{\Omega}_1)\right]=T^{\alpha \beta}=0,
	\label{eq:11}
	\end{array}
	\end{equation}
	where, again, the Weyl contribution is not present as we are considering conformally-flat metric as an ansatz, $\mathcal{O}(\mathcal{C})=0.$ Note that a {\it non-trivial} vacuum solution is given by the vanishing of the Ricci scalar, indeed
	\begin{equation}
	\mathcal{R}=-\frac{6}{F^3}\eta^{\mu\nu}\partial_{\mu}\partial_{\nu}F=-6\frac{2F'+rF''}{rF^3}=0\Longleftrightarrow2F'+rF''=0,
	\label{eq:16}
	\end{equation}
	whose solution for the conformal factor $F(r)$ is:
	\begin{equation}
	F(r)=a-\frac{b}{r},
	\label{eq:17}
	\end{equation}
	where $a$ and $b$ are two integration constants. Note that for generic values of $a$ and $b$ such a solution does not solve the vacuum field equations corresponding to the general quadratic action in Eq.\eqref{eq:1}, but it becomes a solution {\it only} if $a=0$. 
	
	\begin{enumerate}
		\item If $a=0$, we recover the metric solution in Eq.\eqref{ansol} with $n=2$.
		
		\item If $a\neq0$, we can always rescale the coordinates as $x^{\mu}\rightarrow ax^{\mu}$, so that the constant $a$ can be set equal to one without any loss of generality. In this case the conformally-flat metric solution with vanishing Ricci scalar solves the field equations in Eq. \eqref{eq:11} and reads:
		\begin{equation} 
		ds^2=\left(1-\frac{b}{r}\right)^2[-dt^2+dr^2+r^2d\Omega^2].\label{eq:18}
		\end{equation}
		For such a metric the Kretschmann scalar is given by:
		\begin{equation} 
		\mathcal{K}=\frac{8b^2(b^2-4br+6r^2)}{(b-r)^8},\label{eq:18.1}
		\end{equation}
		which seems to be singular for $r=b.$ However, the coefficient $b$ has to be related to the inverse of the scale of non-locality, $b\sim2/M_s$, and since we are working in the regime $M_sr\ll2$, the metric in Eq.\eqref{eq:18} is nothing but the same metric as in Eq.\eqref{ansol}, for $n=2$. Moreover, the corresponding Kretschmann invariant in Eq.\eqref{eq:18.1} is just a constant $8/b^4$, which exactly coincides with the Kretschmann invariant in Eq.\eqref{kret} for $n=2$, if we assume $b=2/M_s$.
		
	\end{enumerate}

	\item {Finally if we set $\mathcal{F}_1(\Box_s)=0$, we will have 

		\begin{equation}
		S= \displaystyle \frac{1}{16\pi G}\int d^4x\sqrt{-g}\left[\mathcal{R}+\alpha_c\left(\mathcal{R}_{\mu\nu}\mathcal{F}_2(\Box_s)\mathcal{R}^{\mu\nu}+\mathcal{C}_{\mu\nu\rho\sigma}\mathcal{F}_{3}(\Box_s)\mathcal{C}^{\mu\nu\rho\sigma}\right)\right].\label{choice 1}
		\end{equation}
		For this action, the vacuum field equations are
		\begin{align}
		P^{\alpha\beta}\approx& \frac{\alpha_c}{{ 8\pi G}} \biggl( 
		4\mathcal{R}_{\mu}^{\alpha}{\cal F}_2(\Box_s)\mathcal{R}^{\mu\beta}
		-g^{\alpha\beta}\mathcal{R}_{\nu}^{\mu}{\cal
			F}_{2}(\Box_s)\mathcal{R}_{\mu}^{\nu}-4\nabla_{\mu}\nabla^{\beta}({\cal
			F}_{2}(\Box_s)\mathcal{R}^{\mu\alpha})
		\\&
		\!\!\!+2\Box({\cal
			F}_{2}(\Box_s)\mathcal{R}^{\alpha\beta})
		\nonumber+2g^{\alpha\beta}\nabla_{\mu}\nabla_{
			\nu}({\cal F}_{2}(\Box_s)\mathcal{R}^{\mu\nu})
		-2\Omega_{2}^{\alpha\beta}+g^{\alpha\beta}(\Omega_{2\sigma}^{\;\sigma}+\bar{
			\Omega}_{2}) -4\Delta_{2}^{\alpha\beta}
		\biggr)\\
	    =& T^{\al\bt}=0\,,
		\label{EOM1}
		\end{align}
		where $\mathcal{O}(\mathcal{C})=0$ if we demand conformally flat metric solutions. One reasonable possibility
		is to investigate solutions such that $\mathcal{R}^{\mu\nu}=0$. Under such constraint, 
		we have to solve the following system of differential equations,
		\begin{align}
		\mathcal{R}_{00}=0\Longleftrightarrow r F(r) F''(r)+r F'(r)^2+2 F(r) F'(r)=0\;,\\
		\mathcal{R}_{11}=0\Longleftrightarrow 3 r F'(r)^2-F(r) \left(3 r F''(r)+2 F'(r)\right)=0\;, \\
		\mathcal{R}_{22}=0\Longleftrightarrow r F(r) F''(r)+r F'(r)^2+4 F(r) F'(r)=0 \;,\\
		\mathcal{R}_{33}=0\Longleftrightarrow r F'(r)^2+F(r) \left(r F''(r)+4 F'(r)\right)=0\;.
		\end{align}		
		Subtracting the first equation from the third one implies the simpler relation $2 F(r) F'(r)=0$. At this stage, we can already 
		restrict solutions to $F(r)=0$ or $F'(r)=0$. If we take $F'(r)=0$, for any $r$, then we are left with $F(r)$ being a constant. We can conclude that in order to have solutions with $\mathcal{R}^{\mu\nu}=0$, the conformally metric ansatz Eq.\eqref{eq:14} should have $F(r)= \rm{constant}$.
		
		We can also search for solutions where generically $\mathcal{R}^{\mu\nu}\neq 0$. In that case, we must find non-trivial vacuum solutions for Eq.\eqref{EOM1}. Note that for the class of possible solutions with $\mathcal{R}=0$, the aforementioned solution Eq.\eqref{eq:18} with $a=0$ is also a solution here. The reason is straighforward, since having $\mathcal{R}=0$ is equivalent to have the form factor $\mathcal{F}_1(\Box_s)=0$ (already discussed above), and since we are looking for conformally flat metrics, the Weyl part of Eq.\eqref{EOM} will be zero. Therefore, solutions with $\mathcal{R}^{\mu\nu}\neq 0$ and $\mathcal{R}=0$ will match those obtained in the previous section. Finally, without additional insight, finding non trivial solutions with $\mathcal{R}\neq 0$, will be a harder task just by noticing the complexity of Eq.\eqref{EOM1}.}	
	
\end{itemize}



\begin{thebibliography}{1}



\bibitem{-C.-M.}C. M. Will, Living Rev. Rel. 17, 4
(2014) {[}arXiv:1403.7377 {[}gr-qc{]}{]}.

\bibitem{-B.-P.}B. P. Abbott et al. {[}LIGO Scientific
and Virgo Collaborations{]}, Phys. Rev. Lett. 116 (2016) no.6, 061102.

\bibitem{-D.-J.}D. J. Kapner, T. S. Cook, E. G. Adelberger,
J. H. Gundlach, B. R. Heckel, C. D. Hoyle and H. E. Swanson, Phys.
Rev. Lett. \textbf{98} (2007) 021101.

\bibitem{-K.-S.}K. S. Stelle, Phys. Rev. D \textbf{16} (1977)
953.

\bibitem{krasnikov} N. V. Krasnikov, “Nonlocal Gauge Theories”, Theor. Math. Phys. 73, 1184 (1987) [Teor. Mat. Fiz. 73, 235 (1987)].

\bibitem{kuzmin}Y. V. Kuz’min, “Finite nonlocal gravity (in Russian),” Sov. J. Nucl. Phys. 50, 1011 (1989) [Yad. Fiz. 50, 1630 (1989)].

\bibitem{Tomboulis:1997gg} 
E.~T.~Tomboulis,
``Superrenormalizable gauge and gravitational theories,''
hep-th/9702146.
	
\bibitem{Tseytlin:1995uq} 
  A.~A.~Tseytlin,
  ``On singularities of spherically symmetric backgrounds in string theory,''
  Phys.\ Lett.\ B {\bf 363}, 223 (1995),
  [hep-th/9509050].

\bibitem{Siegel:2003vt}
  W.~Siegel,
  ``Stringy gravity at short distances,''
  hep-th/0309093.
	
	
\bibitem{Biswas:2005qr} 
T.~Biswas, A.~Mazumdar and W.~Siegel,
``Bouncing universes in string-inspired gravity,''
JCAP {\bf 0603}, 009 (2006)
[hep-th/0508194].
	
\bibitem{Biswas:2011ar} 
T.~Biswas, E.~Gerwick, T.~Koivisto and A.~Mazumdar,
``Towards singularity and ghost free theories of gravity,''
Phys.\ Rev.\ Lett.\  {\bf 108}, 031101 (2012).

  \bibitem{Biswas:2016etb} 
  T.~Biswas, A.~S.~Koshelev and A.~Mazumdar,
  ``Gravitational theories with stable (anti)de Sitter backgrounds,''
  Fundam.\ Theor.\ Phys.\  {\bf 183}, 97 (2016)
  [arXiv:1602.08475 [hep-th]].
  
\bibitem{Biswas:2016egy} 
  T.~Biswas, A.~S.~Koshelev and A.~Mazumdar,
  ``Consistent higher derivative gravitational theories with stable de Sitter and anti de Sitter backgrounds,''
  Phys.\ Rev.\ D {\bf 95}, no. 4, 043533 (2017)
  [arXiv:1606.01250 [gr-qc]].



\bibitem{Koshelev:2017bxd} 
A.~S.~Koshelev and A.~Mazumdar,
``Do massive compact objects without event horizon exist in infinite derivative gravity?,''
Phys.\ Rev.\ D {\bf 96}, no. 8, 084069 (2017)
[arXiv:1707.00273 [gr-qc]].

\bibitem{Frolov:2015bia} 
  V.~P.~Frolov, A.~Zelnikov and T.~de Paula Netto,
  ``Spherical collapse of small masses in the ghost-free gravity,''
  JHEP {\bf 1506}, 107 (2015)
  [arXiv:1504.00412 [hep-th]].
  
  \bibitem{Frolov:2015usa} 
  V.~P.~Frolov and A.~Zelnikov,
  ``Head-on collision of ultrarelativistic particles in ghost-free theories of gravity,''
  Phys.\ Rev.\ D {\bf 93}, no. 6, 064048 (2016)
  [arXiv:1509.03336 [hep-th]].


\bibitem{Frolov}V. P. Frolov,  Phys. Rev. Lett. 115, no. 5, 051102 (2015).

\bibitem{Buoninfante:2018xiw} 
L.~Buoninfante, A.~S.~Koshelev, G.~Lambiase and A.~Mazumdar,
``Classical properties of non-local, ghost- and singularity-free gravity,''
arXiv:1802.00399 [gr-qc].


\bibitem{Cornell:2017irh} 
  A.~S.~Cornell, G.~Harmsen, G.~Lambiase and A.~Mazumdar,
  ``Rotating metric in Non-Singular Infinite Derivative Theories of Gravity,''
  arXiv:1710.02162 [gr-qc].

\bibitem{Boos:2018bxf} 
  J.~Boos, V.~P.~Frolov and A.~Zelnikov,
  ``The gravitational field of static p-branes in linearized ghost-free gravity,''
  arXiv:1802.09573 [gr-qc].


\bibitem{Biswas:2010zk} 
  T.~Biswas, T.~Koivisto and A.~Mazumdar,
  ``Towards a resolution of the cosmological singularity in non-local higher derivative theories of gravity,''
  JCAP {\bf 1011}, 008 (2010)
  [arXiv:1005.0590 [hep-th]].

\bibitem{Biswas:2012bp} 
A.~S.~Koshelev and S.~Y.~Vernov,
	``On bouncing solutions in non-local gravity,''
	Phys.\ Part.\ Nucl.\  {\bf 43}, 666 (2012),
	 [arXiv:1202.1289 [hep-th]].
  T.~Biswas, A.~S.~Koshelev, A.~Mazumdar and S.~Y.~Vernov,
  JCAP {\bf 1208}, 024 (2012),
  [arXiv:1206.6374 [astro-ph.CO]].

\bibitem{Tomboulis:2015}	
E.~T.~Tomboulis,
``Nonlocal and quasilocal field theories,''
Phys.\ Rev.\ D {\bf 92}, no. 12, 125037 (2015).


\bibitem{Chin:2018puw} 
  P.~Chin and E.~T.~Tomboulis,
  ``Nonlocal vertices and analyticity: Landau equations and general Cutkosky rule,''
  arXiv:1803.08899 [hep-th].


\bibitem{Talaganis:2014ida} 
S.~Talaganis, T.~Biswas and A.~Mazumdar,
``Towards understanding the ultraviolet behavior of quantum loops in infinite-derivative theories of gravity,''
Class.\ Quant.\ Grav.\  {\bf 32}, no. 21, 215017 (2015).
S.~Talaganis and A.~Mazumdar,
``High-Energy Scatterings in Infinite-Derivative Field Theory and Ghost-Free Gravity,''
Class.\ Quant.\ Grav.\  {\bf 33}, no. 14, 145005 (2016),
[arXiv:1603.03440 [hep-th]].


\bibitem{Biswas:2013cha} 
T.~Biswas, A.~Conroy, A.~S.~Koshelev and A.~Mazumdar,
``Generalized ghost-free quadratic curvature gravity,''
Class.\ Quant.\ Grav.\  {\bf 31}, 015022 (2014),
Erratum: [Class.\ Quant.\ Grav.\  {\bf 31}, 159501 (2014)].
[arXiv:1308.2319 [hep-th]].





\bibitem{Biswas:2013kla} 
T.~Biswas, T.~Koivisto and A.~Mazumdar,
``Nonlocal theories of gravity: the flat space propagator,''
arXiv:1302.0532 [gr-qc].
	

\bibitem{Buoninfante} L. Buoninfante, Master's Thesis (2016),
{[}arXiv:1610.08744v4 {[}gr-qc{]}{]}.



\bibitem{Koshelev:2018hpt} 
A.~Koshelev, J.~Marto and A.~Mazumdar,
``Towards non-singular metric solution in infinite derivative gravity,''
arXiv:1803.00309 [gr-qc].



\bibitem{Koshelev:2018rau} 
A.~S.~Koshelev, J.~Marto and A.~Mazumdar,
``Towards resolution of anisotropic cosmological singularity in infinite derivative gravity,''
arXiv:1803.07072 [gr-qc].

\bibitem{Stelle}  
H. Lu, A. Perkins, C. N. Pope and K. S. Stelle, Phys. Rev. Lett. 114, no. 17, 171601 (2015), [arXiv:1502.01028 [hep-th]]. H. Lu, A. Perkins, C. N. Pope and K. S. Stelle, Phys. Rev. D 92, no. 12, 124019 (2015), [arXiv:1508.00010 [hep-th]]. 


\bibitem{Cardoso:2017cqb} 
  V.~Cardoso and P.~Pani,
  ``Tests for the existence of black holes through gravitational wave echoes,''
  Nat.\ Astron.\  {\bf 1}, no. 9, 586 (2017),
  [arXiv:1709.01525 [gr-qc]].
  V.~Cardoso, S.~Hopper, C.~F.~B.~Macedo, C.~Palenzuela and P.~Pani,
  Phys.\ Rev.\ D {\bf 94}, no. 8, 084031 (2016),
  [arXiv:1608.08637 [gr-qc]].
  
  \bibitem{Buoninfante:2018mre} 
  L.~Buoninfante, G.~Lambiase and A.~Mazumdar, "Ghost-free infinite derivative quantum field theory,"
arXiv:1805.03559 [hep-th].

\bibitem{Mathur:2005zp} 
  S.~D.~Mathur,
  Fortsch.\ Phys.\  {\bf 53}, 793 (2005),
  [hep-th/0502050].
  

\bibitem{Siegel:1988yz} 
  W.~Siegel,
  ``Introduction to string field theory,''
  Adv.\ Ser.\ Math.\ Phys.\  {\bf 8}, 1 (1988)
  [hep-th/0107094].
  W.~Taylor,
  ``String field theory,''
 In *Oriti, D. (ed.): Approaches to quantum gravity* 210-228
  [hep-th/0605202].
  
\bibitem{Frolov-book}
 V.~P.~Frolov and A.~Zelnikov, "Introduction to blackhole physics"
DOI:10.1093/acprof:oso/9780199692293.001.0001


\bibitem{Edholm:2016hbt} 
  J.~Edholm, A.~S.~Koshelev and A.~Mazumdar,
  Phys.\ Rev.\ D {\bf 94}, no. 10, 104033 (2016)
  doi:10.1103/PhysRevD.94.104033
  [arXiv:1604.01989 [gr-qc]].
  
  
 \bibitem{Balasin:2006cg} 
  H.~Balasin and D.~Grumiller,
  ``Non-Newtonian behavior in weak field general relativity for extended rotating sources,''
  Int.\ J.\ Mod.\ Phys.\ D {\bf 17}, 475 (2008)
  [astro-ph/0602519].
 
  
\bibitem{Vladimirov}
V.S. Vladimirov,, "Generalized functions in mathematical physics", Moscow, Izdatel'stvo Nauka, 1976. 280 p. 
  
 \bibitem{Gelfand} 
I. M. Gel'fand and G. E. Shilov, "Generalized Functions", Volume 1, Publication Year: 1964, ISBN-10: 1-4704-2658-7, ISBN-13: 978-1-4704-2658-3.

\bibitem{Koshelev:2017tvv} 
A.~S.~Koshelev, K.~Sravan Kumar and A.~A.~Starobinsky,
JHEP {\bf 1803}, 071 (2018)
[arXiv:1711.08864 [hep-th]].


	
	
	
	
	
	
	
	
	
\end{thebibliography}
\end{document}